# Why Did the Robot Cross the Road? A User Study of Explanation in Human-Robot Interaction

Zachary Taschdjian

Georgia Institute of Technology, Atlanta, GA 30332, USA
H2O.ai, Mountain View, CA 94043, USA
ztaschdjian@gatech.edu

**Abstract.** This work documents a pilot user study evaluating the effectiveness of contrastive, causal and example explanations in supporting human understanding of AI in a hypothetical commonplace human-robot interaction (HRI) scenario. In doing so, this work situates "explainable AI" (XAI) in the context of the social sciences and suggests that HRI explanations are improved when informed by the social sciences.

## 1 Introduction

The relevance of AI in our society is self-evident. With relevance comes ubiquity and with ubiquity come magnified consequences. Couple this with the fact that AI/ML is generally poorly understood by the public and there's a strong possibility of accidental misuse/misinterpretation. Consider the following example. A physicians AI-enabled tool tells her that a patient with pneumonia AND asthma is LESS likely to die than a similar, non-asthmatic patient. This seems counter-intuitive; the presence of asthma suggests a stronger likelihood of a negative outcome. But as Caruana, et. al. found, this is not the case. Asthmatic patients with pneumonia are admitted directly to the intensive care unit and receive aggressive treatment, thus explaining the counter-intuitive finding. (Caruana, et. al., 2015). In this case, a more accurate neural net was rejected in favor of a less accurate rule-based model because physicians could understand how the rule-based model worked. Without explainability this finding could easily be dismissed as model error likely leading to the AI tool not being adopted in mission critical environments like hospitals.[1] We contend that this is fundamentally a problem of education and knowledge acquisition. Blackbox algorithms are not architected with these needs in mind.

Current AI systems are often approached from a technical perspective leading to engineering-centric explanations created by and for AI experts. Despite XAI's popularity, it's extremely easy to get wrong (Kozyrkov, 2018). While explainability, interpretability and causation are relatively new to AI, they are not new concepts. Explanation and causation can be traced back to Aristotle and have been instrumental in philosophy in the intervening millennia. The unmet need for XAI is easily documented for example through DoD and NSF funding opportunities, (Gunning, 2019), the popular press (Kuang, 2017) and in philosophy (Miller, 2018), computer science (Kim, 2015), cognitive psychology (Moreno, et al., 2007) and doubtless others.[2] There is also evidence that the problem of AI model opacity

1 The terms "explainable" and "interpretable" are used interchangeably despite their slightly differing meanings.

2 Each of these references is an example. This is not intended to be a comprehensive list.

(the inverse of XAI) has real social impacts for example in loan origination, pre-trial sentencing, fake news/propaganda and healthcare (as noted above). Could an interdisciplinary approach drawing from these areas lead to more understandable, ethical and effective AI? This study is premised on the answer being "yes".

The study presented here is an attempt to engage these disciplines in the XAI dialogue beyond the theoretical. This work draws on Miller's summary of explanations. Miller relates three types of explanation; contrastive (i.e. "A and not B"), causal (i.e. "A caused B; without A, B would not occur") and examples (i.e. "A is like B, C, D"). (Miller, 2018). Study 2 presented here compares these three types of explanations using three modes of explanation; narrative text, iconographic images and data charts. The contribution of this work is to evaluate the hypothesis that one of these combinations will provide a "better" explanation than others. A possible implication of this is that one or more of these combinations better fits the users mental model given the context of the human-robot interaction in the test.

The remainder of this paper will first outline related prior work in the contributing disciplines and preparatory work. It will then describe the user study. Finally, it will discuss the study and present findings and opportunities for improvement and future work in this area.

### 1.1 Related Work

As noted above, XAI is inherently interdisciplinary. While there's overlap between the disciplines, related work can be organized by discipline for clarity and structure. This is not a comprehensive list of related work but represents the breadth of issues, questions and history of the contributing disciplines.

### 1.2 Prior work in philosophy

Much of the work in philosophy falls into philosophy of mind, formal epistemology and causation but it can be traced back as far as Aristotle's four causes of action (i.e. ways of answering "why" questions) (Reece, 2019). David Hume also did significant work in causation and epistemology some 2000 years later. He believed that just because two events seem to be conjoined there's no evidence that one causes another. In short, that no amount of data ever justifies belief. This leads Hume to dismiss all inductive inference, leading to his famous skepticism that we can ever truly know anything. Others have built on Hume's approach in more pragmatic ways. Popper, for example, suggests an approach built on refutation of theories. (Parusniková, 2019) More recently, researchers have looked at ways of axiomatizing epistemology in the context of AI systems (Vasconcelos, et al., 2018), causal discovery algorithms (Malinsky, et al., 2017) and structural equation modeling approaches (Halpern, et al., 2005), among others. It's important to note that causation is not the same as explanation although they are related.

### 1.3 Prior work in computer science

Computer science has had an obvious fascination with XAI. Explanation was at the heart of Seymour Papert's Learning and Epistemology group at the MIT Media Lab for example. (Papert, 1988). Some early interest in XAI came out of work in machine translation systems and later expert systems.4 XAI has had a renaissance in the last ~5 years. This takes many directions ranging from arguments for inherently interpretable models (Rudin, 2019) to a more pragmatic understanding of explainability (Paez, 2019) to using XAI for model debugging (Winikoff, 2017). There has also been some relevant work in HCI. Grudin for example draws a clear connection between AI and HCI as early as 2009 (Grudin, 2009) and more recently Google (among others) has produced guidelines for HCI practitioners working with AI. (Google, n.d.). Springer, et al take on model explainability and black box algorithms directly demonstrating that users put unfounded trust in systems perceived to be “intelligent”. (Springer, et al., 2017)

### 1.4 Prior work in cognitive psychology

Cognitive psychology has perhaps the most direct relevance to the educational aspects of XAI. A great deal of research has gone into theories of knowledge acquisition and construction. One of the most transformational was Piaget’s application of Maria Montessori’s work to develop the concept of “constructivism”. This empirical approach to pragmatic knowledge acquisition (and accommodation) has had deep impact not just on developmental psychology but on a range of other disciplines. Li et al., operationalized constructivism in the context of predictive analytics. (Li et al., 2017). While Piaget’s work was foundational, Gopnick has contributed of more recent work building on Piaget’s ideas including a useful survey of post-Piaget developments. (Gopnick, 1996). Another relevant concept from the cognitive psych field is mental models; the idea that humans represent knowledge as mental representations such as scripts. Mental models are relevant to the ed tech agenda because their construction and transference are core to understanding in AI. Johnson-Laird did much of the formative work in this area. (Johnson-Laird, 2010). Jonnassen (among others) also made important contributions around mental models in the context of computer supported collaborative learning. (Jonnassen, 1995).

## 2 Methodology

### 2.1 Preparatory Work with Data Scientists

Preparatory work included a series of between-subjects, think-aloud interviews with professional data scientists and developers. Interviews were conducted with developers and data scientists who are colleagues of the author at H2O.ai. The goal of this work was to identify industry best-practices, or at least a rough heuristic of the mental model(s) professional data scientists employ when interpreting and explaining AI predictions.

Additionally, the study evaluated whether the tool/user interface frames or constrains explanation selection. An affirmation that tool selection and interface constrain and/or frame explanation selection would imply that the user’s mental model of the problem is similarly framed and/or constrained by the tool which has relevance for both designers and trainers of such systems. This is somewhat akin to bias caused by leading questions; the explanation type (as implemented in a user interface) frames the users approach to the problem. This is captured by the adage that when the only tool one has is a hammer, every problem starts looking like a nail. It seems likely that confirming that tooling effects explanation interpretability should generalize, at least to other kinds of analytical use cases. To be clear, confirming this hypothesis is outside the scope of this work.

## 2.2 Assumptions

This study assumes that the choice of AI tool constrains the kinds of explanations that are selected. For example, if the tool uses static (i.e. non-interactive) charts, then it could be expected that explanations using contrastive/counterfactual "what-if" scenarios would not be possible since they would require re-scoring the model (often a cumbersome and time-consuming operation on large data sets). More concretely, in a use case asking why the model denied a loan to a particular person, the user might cite a model feature called "missed payments" and draw the conclusion that this person had a high number. The professional data scientists involved in this preparatory work noted that more analysis would be required to conclude that the high number in variable "missed payments" *caused* the model to classify this person as a risk or *caused* it to deny the loan. In other words that a lurking variable was equally likely to cause this phenomenon and/or that by Simpsons Paradox, this phenomenon might disappear or reverse on deeper analysis. However, the underlying assumption was confirmed (at least qualitatively) by the participants

## 2.3 Methods

This study was a blind, semi-random, between-subjects study of 120 participants using a Google form administered in 3 cohorts via Amazon Mechanical Turk (AMT). Participants were compensated $0.45 for completing the ~5-minute survey question. Participants could only complete one study one time to avoid priming bias. The only qualifications to participate were English fluency and the ability to access AMT (i.e. have an account and a computer/network capable of connecting). English fluency was somewhat problematic as AMT workers opted-in and not all of their levels of fluency appeared equal. The study was conducted in three phases, each phase corresponding to a row in table 1 (contrastive, causal and example). The test scenario was designed to be easily understood by a non-technical audience. Clearly, this scenario doesn't generalize to more complicated use cases or use cases requiring subject matter expertise. The test scenario was the following:

*"Pretend you have a robot that folds your clothes. The robot has accidentally folded a shirt which is inside out, presumably because it couldn't see the tag. Below are three ways the robot can explain the mistake. Please pick the* ***best*** *explanation."*

Participants were then shown the answer choices in Table 1 corresponding to their cohort. The cohorts were the explanatory modes listed in column 1 of table 1. Users could select explanations based on text, images or chart/visualizations. For each cohort, the selected explanation type (i.e. contrastive, causal and example) was shown using each of the three delivery modes shown in table 1. Text explanations used narrative written text to explain the robots mistake. Icons used simple, iconographic images to illustrate the explanation. Charts used simple data visualizations as an explanatory device. Answers were presented in the same order between cohorts for consistency.

A survey asking the following questions was also included. Question 1 used a freeform text field and question 2 used a 5-point Likert scale;

1) *Why did you select the answer you did*?
2) *On a scale of 1 to 5 (1 is worst, 5 is best) how well did the answer you selected explain the robots mistake*?

Asking about scale was intended to elicit how confident participants were in their answers. For example, someone might select an answer believing it was the best choice among a number of poor choices, in which case they could rate it accordingly (i.e. it answered the question but badly).

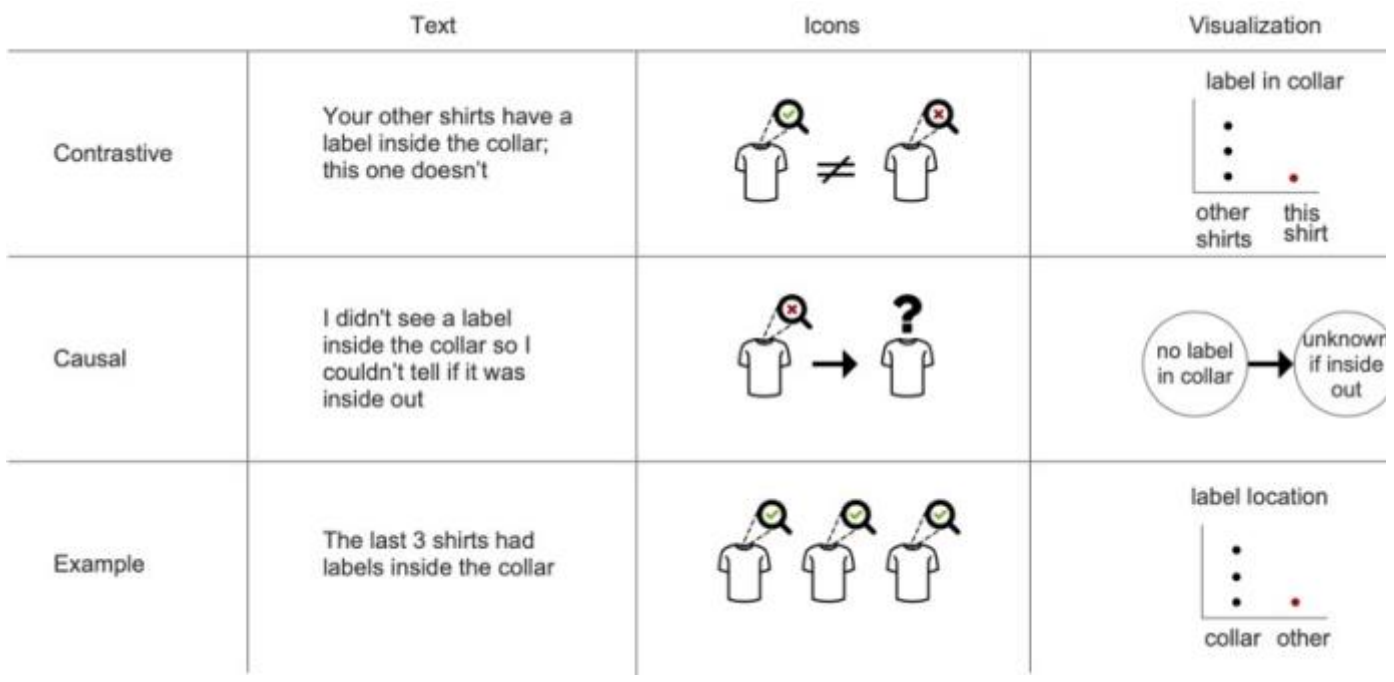

**Table 1.** The matrix of answer selections shown to each participant.

### 2.4 Analysis

The raw answer selections (i.e. answer frequency) for each cohort were analyzed using Python in a Jupyter notebook. Analysis was done following the industry standard sequence for exploratory data analysis of ingestion, cleaning, visualization and analysis. A Chi square was used to analyze correlation.

The question "*why did you select this answer*?" was analyzed separately using NLP techniques for thematic analysis. Tokenization and lemmatization were done using NLTK (wordnet and porter). Unsupervised topic modeling was done with Gensim. Results were displayed based on saliency (relevance metric = 1) and intertopic distance of the principal components using multidimensional scaling. The most salient words were visualized in a word cloud (Table 4).

### 2.5 Results

120 individuals participated; 40 for each explanation cohort. It's hard to draw statistically valid conclusions with such a small sample size, but **participants showed a clear preference for text explanations, especially contrastive explanations** as shown in Table 2 and Figure 1.

| Explanation | Mode | Frequency Scale |
|---|---|---|
| Contrastive | Charts | 4 |
| | Images | 12 |
| | Text | 25 |
| Causal | Charts | 13 |
| | Images | 11 |
| | Text | 16 |
| Example | Charts | 9 |
| | Images | 11 |
| | Text | 21 |

**Table 2.** Results by selection frequency.

Visualizing the results of study 2 showed a clear preference for contrastive, text-based explanations. Explanations using charts had the lowest explanatory power. Participants had a high degree of certainty in their answer selections based on the Likert scale question "how well does your selected answer explain the error?" graphed by frequency. Likert scale; 1 lowest, 5 highest. Almost 50 chose 5 and almost 60 chose 4, the first and second highest levels of confidence, respectively. (see figure 2)

For the NLP analysis of the freeform text field ("why did you choose this answer?"), the top 3 words in the word count were "understand" (19 times), "robot" (16 times) and "explains" (10 times). (see figure 3). However, this demonstrates the obvious fact that there's little useful context in this finding.

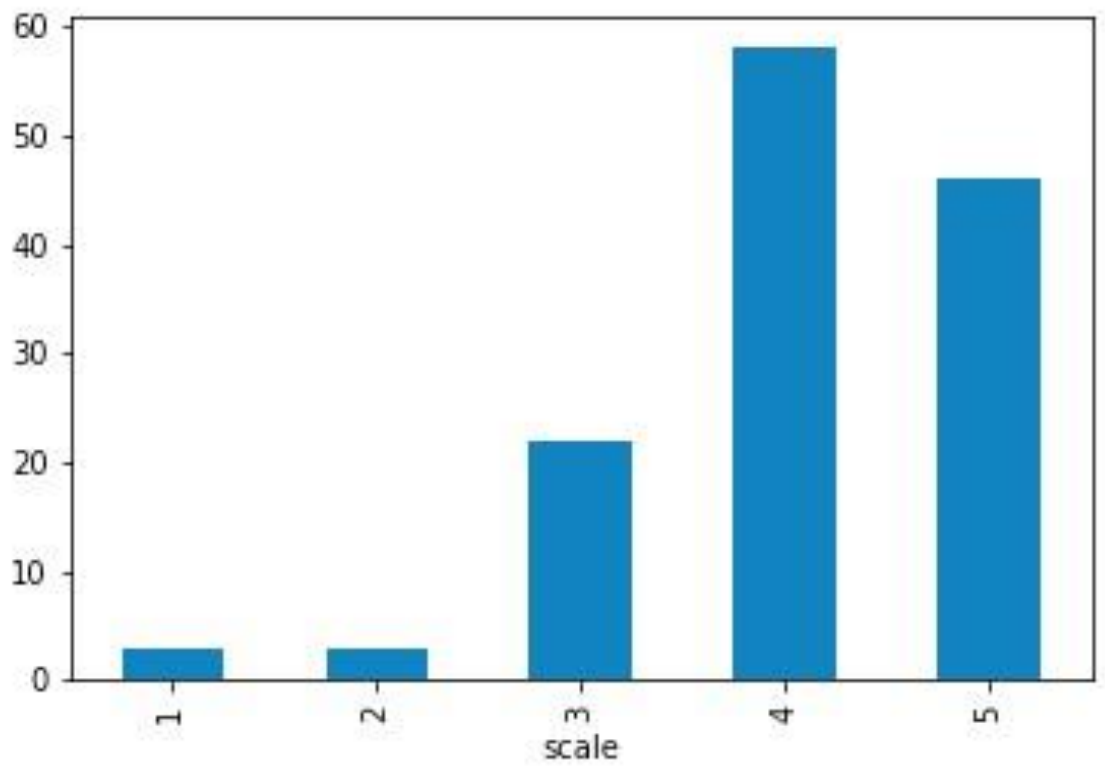

**Table 3.** Frequency of participants response to the question "how well does your selected answer explain

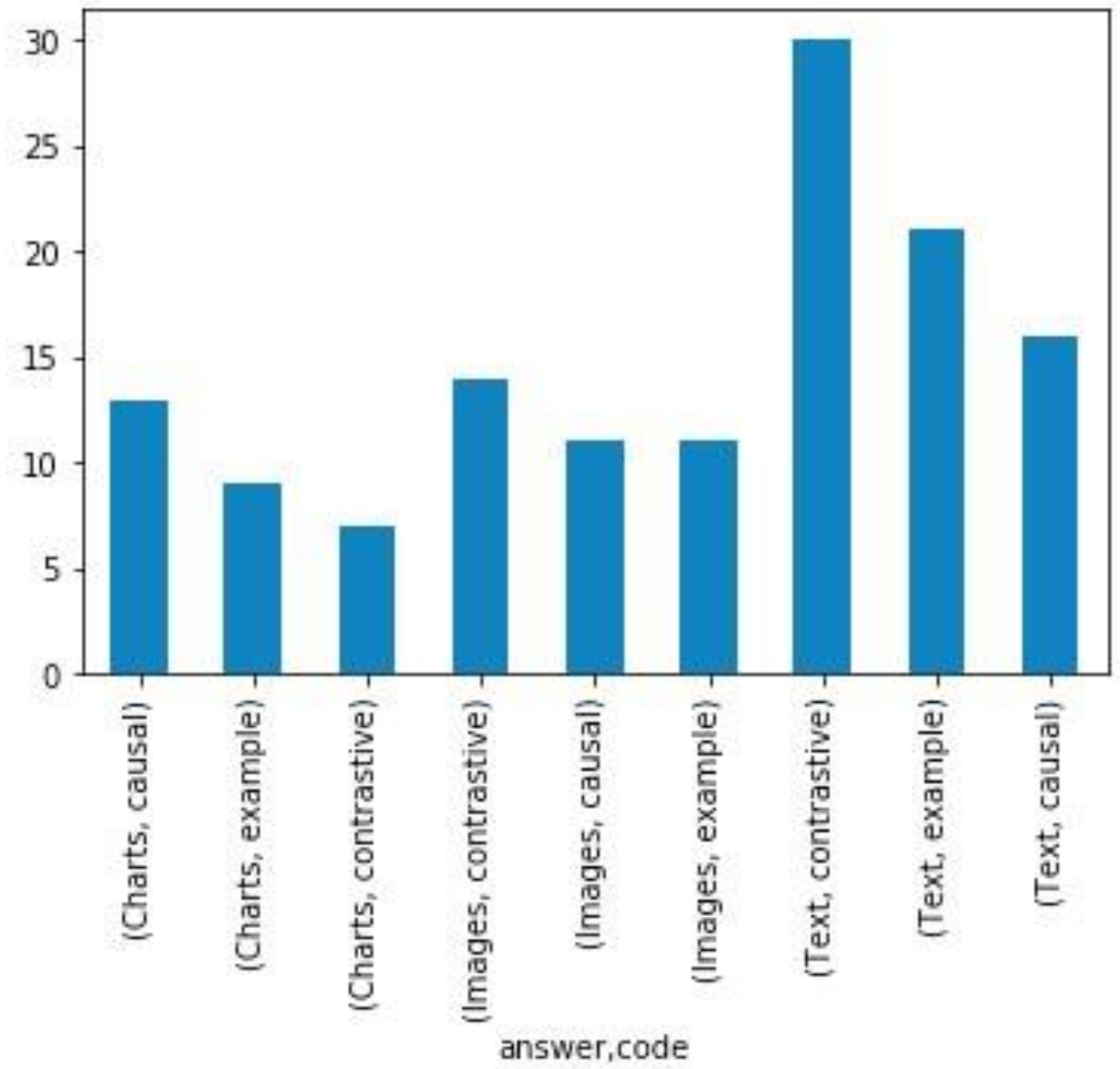

**Table 4.** Chart of results by selection frequency.

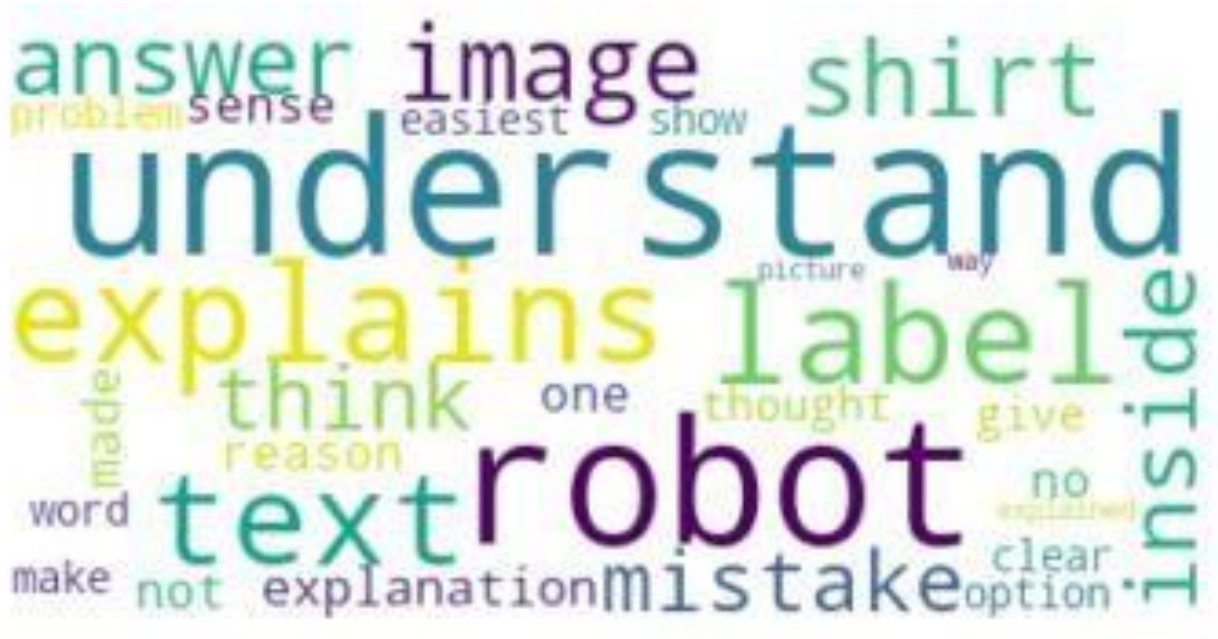

**Table 5.** Word cloud of top 30 terms in the freeform text field. Larger size indicates higher frequency.

# 3 Discussion

## 3.1 Limitations of the Study

As mentioned above, it's difficult to draw statistically valid conclusions from such a small sample. Additionally, there are a number of confounding factors which are difficult to control for. For example, achieving perfect parity between the modes of representation is likely impossible. There are many possible ways to articulate a concept using images, charts and text. Having a perfect mapping between them for the same concept would be difficult. For example, a text explanation could be worded a number of ways, a chart could use lines, bars, scatter or other plotting approaches.

Similarly, more complex scenarios might not map neatly to informational charts or iconographic images. It's also questionable as to whether this strategy would generalize to more complex concepts. The test scenario was chosen because it was perceived to be the most understandable to the widest possible group of AMT workers. While comprehension might be possible with such simple concepts, a more complicated robot error state would be harder to convey. Especially one which required subject matter expertise. Another confounding variable is the inability to easily control for language proficiency in AMT. For example, non-English speakers may be at a disadvantage in answering English language text-based questions. It's impossible to know from the data but given the grammar and spelling choices of some respondents, it's likely that at least a few of them were not native English speakers.

This study could also be simplified by reducing it to a single independent variable (i.e. explanation type), rather than introducing other complicating variables in the delivery mode. The study introduced unnecessary complexity by using larger cohorts. Perhaps a better approach would build cohorts around one explanation type and one delivery mode.

Subsequent studies could also introduce design variations in the images, text and charts. For example, each of these groups could have variations in design/wording to control for subtle discrepancies between them. As noted above, it would also be interesting to run the test with a wider variety of use cases with varying levels of complexity. This would help determine if the results generalize beyond the simplest use cases. Given sufficient time and funding, higher quality data could probably be achieved without using AMT.

### 3.2 Free-form Responses

The NLP analysis of the freeform text is also somewhat problematic. The structure of the question elicited trivialities like "It was the most clear" or "It makes the most sense". Asking users to articulate why they chose one answer over another is likely unrealistic, especially in an AMT study. AMT workers are generally incentivized to complete a task as quickly as possible and don't appear especially vested in the quality of their responses. It was extremely difficult to control for participants who randomly wrote in whatever triviality they came up with.

However, this does point to a possibly deeper observation; that participants are actually unaware (rather than unable to articulate) why they selected one answer over another and hence they resorted to trivialities. That is, the participants are asked to ascribe causes to the robot's behavior based on its internal "mental state". Perhaps in the absence of obvious causal agents, participants frame the robots response using their own unconscious causal scripts as if they were in the robot's situation. This supposition is related to the concept of "unconscious bias" and the idea that our own internal mental states are unavailable to our reflective, conscious mind. This remains outside the scope of this work, but if true, compensating for this phenomenon would require further research in areas such as pre-attentive processing.

## 4 Future Work

This work was intended to evaluate a simple, universal human-robot interaction. Future work might address more complex use cases and use cases requiring subject matter expertise on the part of the human interlocutor. This could involve scenarios designed to discover the limits of visual explanations. For example, would the explanation types generalize to human-robot teaming scenarios like emergency response or on the battlefield? These types of scenarios would introduce many complicating variables such as comprehending robot state under time constraints.

Another promising scenario involves introducing interactivity in the human-robot interaction. Much of the work cited in section 1.4 draws on the concept of "constructionism" in which users construct explanations via interaction with their environment. Introducing interactivity to the explanation scenarios evaluated in this paper would likely elicit a very different set of responses. For example, allowing users to frame foils in contrastive explanations such as “why did you take action A and not action B?”

Humans use a variety of imperfect methods to communicate explanations to each other including text, images and data among others. Perhaps future work will invent new methods of explanation and ways to illustrate causation based on the strengths of machines. Specialized professional domains like scientific data visualization and AutoML applications have broken interesting ground in these areas recently. Currently, the main limitation appears to be the technical feasibility of implementing a robot-operating system (or ML algorithm more generally) capable of articulating its internal states in any human-understandable format.

## 5 Conclusions

This research contributes a preliminary analysis of contrastive, causal and example-based explanations in a hypothetical but potentially common-place human-robot interaction. These explanation methods are evaluated using text, chart/data visualization and simple images. This research is valuable because understanding how humans learn, explain, interpret and develop mental models relating to AI behaviors is crucial to safe, effective and ethical human-AI interaction.

## 6 References

1. Ackermann, E. (2001). Piaget's Constructivism, Papert's Constructionism: What's the difference? 5.
2. Caruana, R., Lou, Y., Gehrke, J., Koch, P., Sturm, M., & Elhadad, N. (2015). Intelligible models for healthcare: Predicting pneumonia risk and hospital
3. 30-day readmission. Proceedings of the 21th ACM SIGKDD International Conference on Knowledge Discovery and Data Mining - KDD '15, 1721–1730. https://doi.org/10.1145/2783258.2788613
4. Chapman, A., Hadfield, M., & Chapman, C. (2015). Qualitative research in healthcare: An introduction to grounded theory using thematic analysis. Journal of the Royal College of Physicians of Edinburgh, 45(3), 201–205. https://doi.org/10.4997/JRCPE.2015.305
5. Gopnik, A. (1996). The Post-Piaget Era. Psychological Science, 7(4), 221–225. https://doi.org/10.1111/j.1467-9280.1996.tb00363.x
6. Grudin, J. (2009). Ai and hci: Two fields divided by a common focus. AI Magazine, 30(4), 48. https://doi.org/10.1609/aimag.v30i4.2271
7. Gunning, D. (2019). DARPA's explainable artificial intelligence (XAI) program. In Proceedings of the 24th International Conference on Intelligent User Interfaces (IUI '19). Association for Computing Machinery, New York, NY, USA, ii. DOI:https://doi.org/10.1145/3301275.3308446
8. Halpern, J. Y. (2005). Causes and explanations: A structural-model approach. Part i: causes. The British Journal for the Philosophy of Science, 56(4), 843–887. https://doi.org/10.1093/bjps/axi147
9. Johnson-Laird, P. N. (2010). Mental models and human reasoning. Proceedings of the National Academy of Sciences, 107(43), 18243. https://doi.org/10.1073/pnas.1012933107
10. Jonassen, D. (1995). Operationalizing mental models: strategies for assessing mental models to support meaningful learning and design supportive learning environments. In The first international conference on Computer support for collaborative learning (CSCL '95). L. Erlbaum Associates Inc., USA, 182–186. DOI:https://doi.org/10.3115/222020.222166
11. Kim, B. (2015). Interactive and interpretable machine learning models for human machine collaboration. MIT.
12. Kozyrkov, C (2018). Explainable AI won't deliver. Here's why. [Online]. Available: https://hackernoon.com/explainable-ai-wont-deliver-here-swhy-6738f54216be. [Accessed 11 October 2019].Joyner, D. A. (2018a). Intelligent Evaluation and Feedback in Support of a Credit-Bearing
13. MOOC. In Proceedings of the 19th International Conference on Artificial Intelligence in Education. London, United Kingdom. Springer.
14. Kuang, C. (2017, November 21). Can a. I. Be taught to explain itself? The New York Times. https://www.nytimes.com/2017/11/21/magazine/can-aibe-taught-to-explain-itself.html
15. Li, X., & Huan, J. (2017). Constructivism Learning: A Learning Paradigm for Transparent Predictive Analytics. Proceedings of the 23rd ACM SIGKDD International Conference on Knowledge Discovery and Data Mining - KDD '17, 285–294. https://doi.org/10.1145/3097983.3097994
16. Malinsky, D., & Danks, D. (2018). Causal discovery algorithms: A practical guide. Philosophy Compass, 13(1), e12470. https://doi.org/10.1111/phc3.12470
17. Miller, T. (2018). Explanation in artificial intelligence: Insights from the social sciences. ArXiv:1706.07269 [Cs]. http://arxiv.org/abs/1706.07269

18. Moreno, R., & Mayer, R. (2007). Interactive multimodal learning environments: Special issue on interactive learning environments: contemporary issues and trends. Educational Psychology Review, 19(3), 309–326. https://doi.org/10.1007/s10648-007-9047-2
19. Páez, A. (2019). The pragmatic turn in explainable artificial intelligence(Xai). Minds and Machines, 29(3), 441–459. https://doi.org/10.1007/s11023-019-09502-w
20. Papert, S. (1988). One AI or Many? Daedalus, 117(1), 1-14. Retrieved February 19, 2020, from www.jstor.org/stable/20025136
21. Parusniková Z. (2019) Popper and Hume: Two Great Skeptics. In: Sassower R., Laor N. (eds) The Impact of Critical Rationalism. Palgrave Macmillan, Cham
22. People + AI Research. (n.d.). Google Design. https://design.google/library/ai/
23. Reece, B. C. (2019). Aristotle's four causes of action. Australasian Journal of Philosophy, 97(2), 213–227. https://doi.org/10.1080/00048402.2018.1482932
24. Rudin, Cynthia. "Stop Explaining Black Box Machine Learning Models for High Stakes Decisions and Use Interpretable Models Instead." ArXiv:1811.10154 [Cs, Stat], Sept. 2019. arXiv.org, http://arxiv.org/abs/1811.10154.
25. Springer, A., Hollis, V., & Whittaker, S. (2017). Dice in the black box: User experiences with an inscrutable algorithm. AAAI Spring Symposium - Technical Report, SS-17-01-, 427–430.
26. Vasconcelos, M., Cardonha, C., & Gonçalves, B. (2018). Modeling epistemological principles for bias mitigation in ai systems: An illustration in hiring decisions. Proceedings of the 2018 AAAI/ACM Conference on AI, Ethics, and Society- AIES '18, 323–329. https://doi.org/10.1145/3278721.3278751
27. Winikoff, M. (2017). Debugging Agent Programs with Why? Questions. In Proceedings of the 16th Conference on Autonomous Agents and Multi-Agent Systems (AAMAS '17). International Foundation for Autonomous Agents and Multiagent Systems, Richland, SC, 251–259.